# H2020-SPACE-ORIONAS
# Miniaturized optical transceivers for high-speed optical inter-satellite links

L. Stampoulidis[a], A. Osman[a], I. Sourikopoulos[a], G. Winzer[b], L. Zimmermann[b], W. Dorward[i], A. Serrano Rodrigo[h], M. Chiesa[h], D. Rotta[h],  A. Maho[c], M. Faugeron[c], M. Sotom[c], F. Caccavale[g]

[a] LEO Space Photonics R&D, Lefkippos Tech. Park, 27 Neapoleos Str., Ag. Paraskevi, 15341, Athens, Greece
[b] IHP GmbH, Frankfurt (Oder), Germany
[c] Thales Alenia Space, 26 av. J-F Champollion, 31037 Toulouse Cedex 1, France
[g] Thales Alenia Space Switzerland AG, Schaffhauserstrasse 580, 8052 Zürich, Switzerland
[h] Fondazione InPhoTec, Scuola Superiore Sant' Anna, Via Giuseppe Moruzzi 1, 56124, Pisa, Italy
[i] ALTER TECHNOLOGY UK, 5 Bain Square, EH54 7DQ Livingston, Scotland, UK

## ABSTRACT

The European H2020-SPACE-ORIONAS project targets the development of optical transceiver and amplifier integrated circuits and modules applicable to high-speed and compact laser communication terminals. This paper presents the most recent project achievements in the fabrication of high-speed electronic-photonic modulator and receiver circuits monolithically integrated in the silicon photonics platform and their assembly in bread-board level photonic modules.

**Keywords:** optical inter-satellite links, satellite constellations, photonic integrated circuits, space photonics

## 1. INTRODUCTION

The sustained entry of optical inter-satellite links (OISLs) in satellite constellations requires the availability of compact photonic devices to deliver transmission and amplification of the high-speed optical signals. Following the trends in terrestrial optical communications where photonic integration has been the tool to continuously shrink the physical size of transceiver modules, modern OISLs are also expected to leverage photonic integrated circuits (PICs) made within semiconductor foundries to combine compactness and cost-effectiveness. Space transceiver PICs are also expected to merge optical and electronic functions into a single chip to optimize footprint and RF performance, but also to relieve the system costs from the procurement and manual assembly of discrete, rad-hard electronic circuits. In this context silicon photonics (SiPh) is of particular interest for the development of OISL transceivers since it allows monolithic integration of optical waveguides, opto-electronics and electronic integrated circuits within a CMOS process. Such transceivers have now reached the stage of pre-flight testing [1]. Moving forward, flavors of SiPh that engage SiGe bipolar transistor technologies to manufacture the transceiver electronic circuitry would be of special interest due to their inherent radiation hardness.

In this paper we present recent advancements of the European H2020-SPACE-ORIONAS project on the development of compact transceiver integrated circuits applicable to new generation 1.55 um, high-speed OISLs. Transceiver circuits were designed and fabricated in IHP 0.25 um photonic-BiCMOS process which enables monolithic integration of active/passive photonics with high-speed SiGe BiCMOS electronics. Mach-Zehnder modulators (MZMs) demonstrate monolithic integration of carrier depletion phase shifters in silicon and segmented BiCMOS driver circuits. Similarly receivers demonstrate monolithic integration of MMI coupler demodulators, Ge diodes and transimpedance BiCMOS circuits. >20 active/passive photonic elements together with the BiCMOS circuits were integrated on a single chip with a footprint of <15 mm$^2$ (MZM) and 5.0 mm$^2$ (receiver) respectively. We demonstrate  the feasibility of edge-coupled bread-board level assembly of the integrated circuits with DFB laser diodes (transmitter sub-assembly) and fiber arrays (receiver assembly) enabled by on-chip integrated spot size converters (SSCs).

## 2. OISL SPECIFICATION

The ORIONAS optical system architecture has been presented in [3]. In summary, the optical link configuration in terms of operating wavelength, data rate and distance is as follows:

This work was supported by the H2020-SPACE-ORIONAS project from the European Union's Horizon 2020 research and innovation program under grant agreement No. 822002. This is a draft manuscript submitted to SPIE for publication. Posting of draft manuscript is permitted under SPIE sharing policies: https://www.spiedigitallibrary.org/article-sharing-policies

- Operating wavelength is 1.55 µm, C-band,
- Data-rate is 20 – 50 Gb/s
- Full-duplex/bidirectional capabilities are required with symmetric throughput,
- Target distance is spanning from 1000 to 6000 km depending on link type (intra- vs. inter-plane) and constellation geometry.

The table below shows an exemplary link budget. The specified modulation format is DPSK and as such receiver integrated circuits must accommodate balanced detection. The optical pre-amplifier should be able to handle input powers below -40 dBm.

| Link parameter | Level | Unit | Note |
| --- | --- | --- | --- |
| Optical launch power | +32 | dBm | |
| Power sharing | 0 | dB | 1 channel |
| Tx telescope path losses | -4 | dB | including obscuration |
| Radiated power/channel | +28 | dBm | |
| Tx telescope gain | 107.7 | dB | 12 cm aperture diameter |
| Tx depointing losses | -1.7 | dB | @ +/- 5 µradians |
| Free-space losses (dB) | -273.7 | dB | @ 6000 km distance |
| Rx telescope gain | 107.7 | dB | 12 cm aperture diameter |
| Received power/channel | -32 | dBm | |
| Rx telescope path losses | -3 | dB | including obscuration |
| Fiber coupling loss | -4 | dB | single-mode fiber coupling |
| Fiber coupled power | -39 | dBm | |
| Rx sensitivity ($10^{-4}$ BER) | -41 | dBm | optically-preamp. DPSK @25 Gb/s ($10^{-3}$ BER) |
| Link margin | 2 | dB | |

## 3. INTEGRATED OPTICAL TRANSCEIVERS

### 3.1 Transmitter PIC and sub-assembly

The figure below shows the MZM PIC fabricated in IHP SG25EPIC process. Signal modulation relies on the plasma dispersion effect, namely electrical modulations of free carriers within the silicon waveguide. The PIC hosts two MZMs (only one MZM was used for the transmitter sub-assembly), having 16 phase shifter elements and one heater structure in each arm. Each phase shifter section is implemented as a pn doped waveguide, to be operated in carrier depletion mode, driven by a dedicated driver circuit. The heater section is a spiral waveguide equipped with a metal line close by, acting as a resistive heater and used to fine-tune the MZM operation. The chip is layout so that RF interfaces are grouped into one side of the PIC while the opposite side is assigned to the optical interfaces. This configuration allows assembly with external optical devices such as laser diodes as well as coupling of optical I/O using fiber array components. Optical I/O is realized through on-chip SSCs which feature a 127 um pitch, and exhibit an insertion loss to standard SMF fiber <1.5 dB across the C-band. Optical I/Os include input/output interfaces to the MZMs as well as two additional ports which are connected through an integrated loop within the chip and used for optical alignment during assembly with external optics. DC pads are distributed along the horizontal side of the PIC providing access to the heater elements as well as providing ground and supply voltages of the driver stages and driver bias current control.

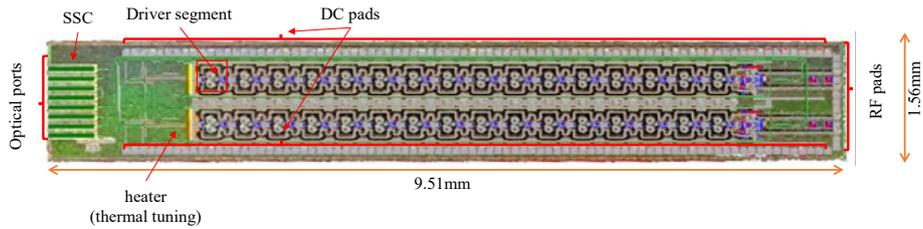

Figure 1. Fabricated MZM SiPh PIC with layout overlay

The opto-electronic performance estimated by layout simulations is as follows: a) Vπ*L = 2.9 V*cm, b) ER = 11 dB, c) RF output swing (differential) at the modulator 4 V, d) RF input swing (differential) ranges from 300mV - 800mV, e) Differential input impedance 100 Ω (GSSG configuration), f) Heater resistance: 100 Ohm - Tuning power: 26 mW / pi, g) Power consumption per channel = 1.6 W.

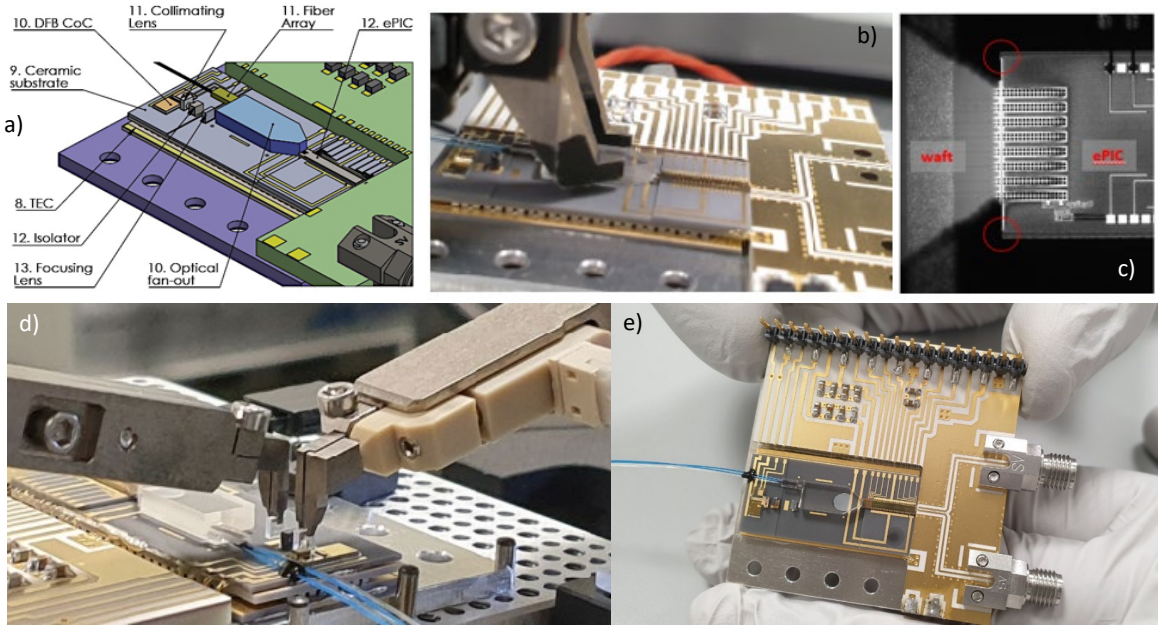

Figure 2. Transmitter optical sub-assembly: a) CAD model, b) WAFT to PIC alignment, c) detail of coupling area between WAFT and EPIC, d) micro-optic assembly process and e) finished sub-assembly

In order to verify the feasibility of edge coupling with external laser diodes and fiber arrays, a bread-board transmitter optical sub-assembly (TOSA) was fabricated. Figure 2a) shows the TOSA design which relies on the use of an optical fan-out as an interface between the MZM PIC and the external active/passive optical devices. The optical fan-out is a glass waveguide array to fiber transposer (WAFT) manufactured by an ion-exchange process and used to: a) increase the pitch of the EPIC optical outputs (127 µm) and separate the input port of the MZM to allow coupling to free space optics and a DFB laser diode and b) further expand the spot size at the PIC output to optimize coupling to standard SMF fiber. As shown in the figures above, the WAFT comes with a pre-assembled fiber array that provides the fiber interfaces to the chip output and alignment waveguides. The insertion losses of the fan-out are <0.7 dB for the non-fibered and <1.2 dB for the fibered channels. Figure 2c) shows the detail of the coupling region between WAFT and PIC. By using the alignment loop to launch and collect light from an external laser, the insertion loss between WAFT and PIC is monitored in one of the fiber ports of the WAFT fiber array. The minimum measured insertion loss between the two ports of the fiber array was 5 dB – this is in agreement with the theoretical expected value, which is broken down as follows: 3 dB attributed to 2x SSCs and <2.4 dB attributed to 2x WAFT fibered waveguides.

Figure 2d) shows the assembly process of the coupling micro-optics used between the PIC and an edge emitter laser diode. The micro-optic assembly consists of a first lens that collimates the diverging laser beam, an isolator to prevent back-reflections and a second lens that focuses the beam to the WAFT input optical waveguide. The nominal coupling loss of the sub-assembly is estimated at 1.42 dB. Figure 2e) shows the finished assembly. The complete assembly sits on top of a ceramic AlN substrate which is attached to a TEC device. Access to the MZM RF ports have been facilitated by wire-bonding of the MZM RF pads to an external PCB which also provides routing and access to the DC signals. Device testing is in progress.

### 3.2 Receiver PIC and sub-assembly

Figure 3a) shows the block diagram of the twin channel receiver PIC fabricated in the same process.

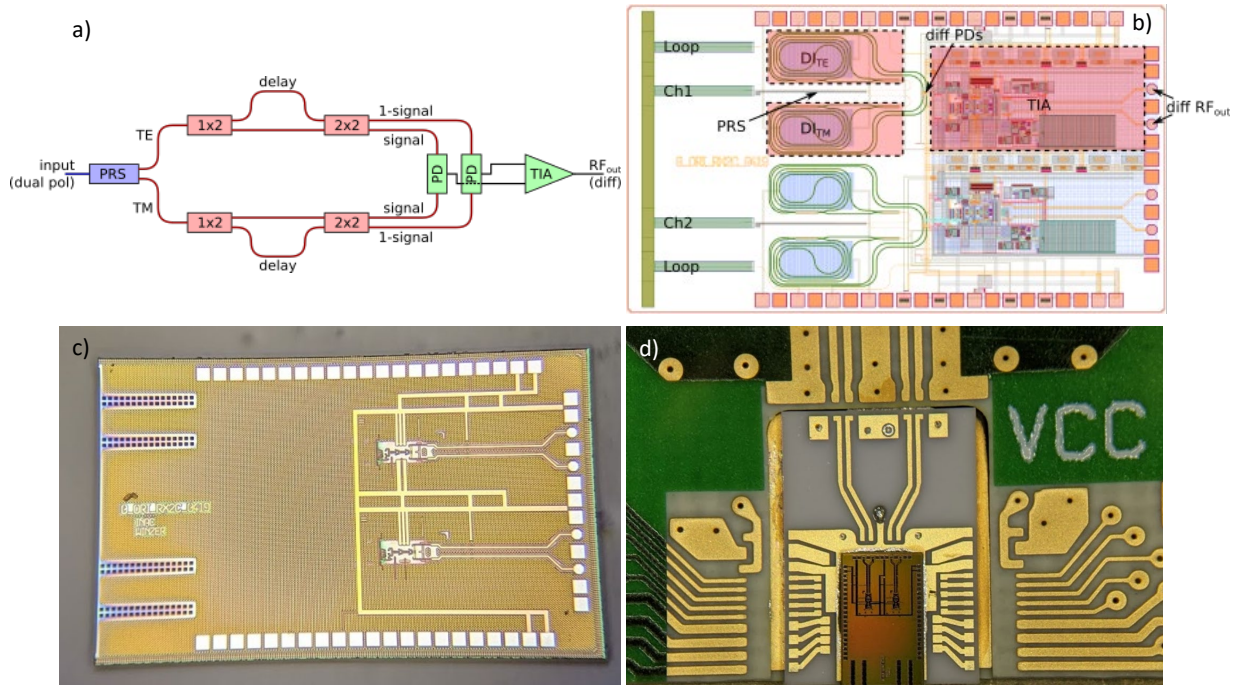

Figure 3. SiPh Receiver PIC and sub-assembly: a) block diagram of receiver circuit, b) layout overview, c) fabricated integrated circuit and d) PIC bonded on ceramic carrier.

The PIC layout is shown in figure 3b) with the demodulator and transimpedance photonic and electronic building blocks marked in the dashed areas. The chip includes two channels, each one having one optical input interface. Similar to the MZM chip, there are two additional optical inputs (marked as "loop") which can be used for alignment during pigtailing. DPSK modulated light is launched into the chip via the integrated SSCs and split to its TE and TM polarization signal components in an integrated Polarization Rotator Splitter (PRS). For the demodulation a Delay Interferometer (DI) is implemented. This interferometer consists of a power splitter (1×2), a combiner (2×2) and two optical path in between, while the optical path length of one of both is exactly one bit length longer. Due to the optical path length difference of one bit between the arms, the interference in the power combiner is defined by the phase difference of two adjacent bits. Due to the built-in phase relation of the power combiner, one of the outputs carries the interference pattern and the other one the inverted one (1-signal). The signals of both outputs are received by two photodiodes (one for each output), which are connected to the input stage of the Transimpedance Amplifier (TIA). The optical path for the second polarization is exactly the same – just mirrored. The receiver is designed as polarization insensitive. Therefore it can be assumed that the bit pattern in both polarizations are aligned and carry the same information. This allows to use the same set of photodiodes for the reception of both polarizations. Photodetection is done by a pair of Ge diodes waveguide coupled to the DIs. The process design kit includes Ge diodes with bandwidth of 40 GHz this surpasses the bandwidth requirements of the OISL target data rate of 20 – 50 Gb/s. The outputs of the photodiodes is connected to the TIA circuit which is used to convert the small input currents from the photodiodes into a differential output voltage. A cascaded 3-stage amplifier circuit provides the amplified differential output RF signal. The photo-generated current from the detectors is converted to voltage by the input stage and amplified to a sufficient voltage swing by the gain stage The third output buffer stage enables to drive external differential 100-Ω load.

The diced receiver PIC is shown in figure 3c) and the chip bonded on a ceramic carrier ready for assembly is shown in figure d). The PIC hosts >20 opto-electronic elements (excluding the BiCMOS circuitry) into a chip areas of <5 mm$^2$.

## 4. CONCLUSIONS

We have reported the latest R&D activities of H2020-SPACE-ORIONAS project in the field of 1.55 um OISLs. We presented fabrication of electronic-photonic modulator and receiver integrated circuits in IHP photonic BiCMOS process and their bread-board assembly in transmitter and receiver optical sub-assemblies. The project continues with the assembly and system-level testing of these components into the end-user transmission testbed.

## 5. ACKNOWLEDGMENT

The H2020-SPACE-ORIONAS project has received funding from the European Union's Horizon 2020 research and innovation programme under grant agreement No. 822002.